\documentclass[english,xaps,prb,twocolumn]{revtex4}
\usepackage{amsmath}
\usepackage{amssymb}
\usepackage{graphicx}

\makeatletter
\@ifundefined{textcolor}{}
{%
 \definecolor{BLACK}{gray}{0}
 \definecolor{WHITE}{gray}{1}
 \definecolor{RED}{rgb}{1,0,0}
 \definecolor{GREEN}{rgb}{0,1,0}
 \definecolor{BLUE}{rgb}{0,0,1}
 \definecolor{CYAN}{cmyk}{1,0,0,0}
 \definecolor{MAGENTA}{cmyk}{0,1,0,0}
 \definecolor{YELLOW}{cmyk}{0,0,1,0}
}

 \usepackage{amsbsy}
\usepackage{mciteplus}
\usepackage{babel}

\makeatother

\usepackage{babel}
\begin{document}

\title{Wave function description of conductance mapping for quantum Hall electron interferometer}

\author{K. Kolasi\'nski}
\affiliation{AGH University of Science and Technology, Faculty of Physics and Applied Computer Science,\\
al. Mickiewicza 30, 30-059 Krak\'ow, Poland}
\author{B. Szafran}
\affiliation{AGH University of Science and Technology, Faculty of Physics and Applied Computer Science,\\
al. Mickiewicza 30, 30-059 Krak\'ow, Poland}

\date{\today}

\begin{abstract}
Scanning gate microscopy of quantum point contacts (QPC) in the integer quantum Hall regime is considered
in terms of the scattering wave functions with a finite-difference
implementation of the quantum transmitting boundary approach.
Conductance ($G$) maps for a clean QPC as well as for a system including an antidot within
the constriction are evaluated. The step-like locally flat $G$ maps for clean QPCs
turn into circular resonances that are reentrant in external magnetic field when the antidot is introduced to the constriction.
 The current circulation around the antidot and the spacing of the resonances
at the magnetic field scale react to the probe approaching the QPC. The calculated $G$ maps with a rigid but soft antidot potential
reproduce the features detected recently in the electron interferometer [F. Martins et al. Nature Sci. Rep. {\bf 3}, 1416 (2013)].
\end{abstract}

\maketitle

\section{Introduction}
Transport properties of devices based on two-dimensional electron gas in the integer quantum Hall regime
are determined by
uncompensated currents that are carried by the edges of the sample.\cite{Martin}
The edge states have a very large coherence length \cite{Martin} and they can be used for construction of electron interferometers.\cite{halperin82,alphenaar92}
Each of the edges carry the current in a single direction only and
the backscattering \cite{bu} at high magnetic field requires electron transfer from one edge to the other across the bulk of the sample.
The interedge tunneling paths can be opened at constrictions (quantum point contacts QPCs) which are intentionally introduced to the channel. A pair of QPCs
with an internal cavity \cite{ku1} form a setup which is referred to as quantum Hall,\cite{rosenow,hackens,qh2} electronic Fabry-P\'erot \cite{fphalperin,fpcz,McClure} or Aharonov-Bohm  (AB) interferometer.\cite{zozu1,caminoprb,caminoprl}
The latter reference is due to the periodicity of conductance ($G$) in external magnetic field for the currents following the edges of the cavity.
A similar interference mechanism and periodic $G$ behavior is found for an antidot introduced between the edges of the sample\cite{zozuantidot,hackens,anti1,anti2,anti3,anti4,anti5,anticz}
with electron currents encircling the antidot.

At low magnetic field the electron currents passing through a quantum point contact can be mapped\cite{qpc0,qpc1,qpc2,qpc4,qpc8}  by
the scanning gate microscopy \cite{sgmr}  which measures conductance as functions of the position of the atomic force microscope tip moving above the sample.
The charged tip is capacitively coupled to the electron gas and modifies the local potential landscape.
In particular a clear semi-classical magnetic focusing \cite{wff} of electron currents at a field of a fraction
of Tesla was observed. The magnetic fields of the order of mT lift the interference pattern of $G$ maps of QPC that appear
according to the weak localization mechanism.\cite{paradiso}
At higher magnetic fields, within the quantum Hall regime the currents evade a direct mapping bypassing any potential perturbations introduced by the tip.
The conductance in the quantum Hall regime can still be affected by the tip when it enhances the inter-edge tunnel coupling,\cite{edge1,edge3}
depopulate the edge states within QPC,\cite{edge2} or allow for selective control of individual edge channels.\cite{paradiso}

Scanning gate microscopy \cite{hackens,qh2} was used for detection of the charging effects \cite{fpcz,edgc,qpc6cz} of the Coulomb island in the interferometer including an intentionally
introduced antidot. Recently,\cite{martinsglowna} a spontaneous formation of an interferometer
with a quantum Hall island (QHI) located inside a quantum point contact was demonstrated by the scanning gate microscopy. Simulations
of the coherent transport in similar conditions are the purpose of the present work.
To the best of our knowledge we provide the first wave function description of the scanning gate microscopy mapping of the coherent flow across the
electron interferometer. The QHI is modeled as an antidot with a fixed potential. We discuss formation of current loops around the QHI which is reentrant
in function of the magnetic field. We describe perturbation to the current flow pattern introduced by the tip and the
consequences of bridging the edge currents by the tip for conductance.
The present numerical simulations reproduce the step-like character of experimental \cite{edge2,paradiso,paradiso2} integer quantum Hall $G$ maps for clean QPCs
with flat minima near the QPCs and no distinct features for the tip outside of the QPC. Calculations for the electron interferometer
reproduce the characteristics of experimental SGM maps,\cite{martinsglowna} including the circular form of oscillations in the $G$ map, the shifts of resonant lines to lower values of $B$ by the repulsive
tip as well as the reduction of the $G$ periodicity for the tip approaching the QHI. We demonstrate that the latter occurs only when the potential
of the QHI potential has a soft profile.

Below we discuss the periodicity of the conductance oscillations.
The early experiments \cite{ku1,anti1} on electron interferometers in the integer quantum Hall regime detected the Aharonov-Bohm periodicity with period $\Delta B=\Phi_0/S$, where
$\Phi_0=e/h$ is the flux quantum and $S$ is the area encircled by the currents. Subsequent studies \cite{caminoprl,goldman14} reported
 fractional periodicity with  $\Delta B=\Phi_0/(f_c S)$, where $f_c$ is the number of edge modes fully transmitted across the sample.
The fractional periodicity of AB conductance oscillations in electron interferometers are explained as due to the electron-electron interaction.\cite{rosenow}
The interaction effects leading to fractional periodicity are outside the range of mean field description\cite{zozu1,zozuantidot} which reproduces the $\Delta B=\Phi_0/S$ period.
The present calculation neglects the electron-electron interaction and in consequence the integer periodicity is found for any $f_c$.
We focus on the qualitative changes of the AB period that are due to the presence of the tip which are
independent of $f_c$.

\section{Model}

\begin{figure*}
\begin{centering}
\includegraphics[width=10cm]{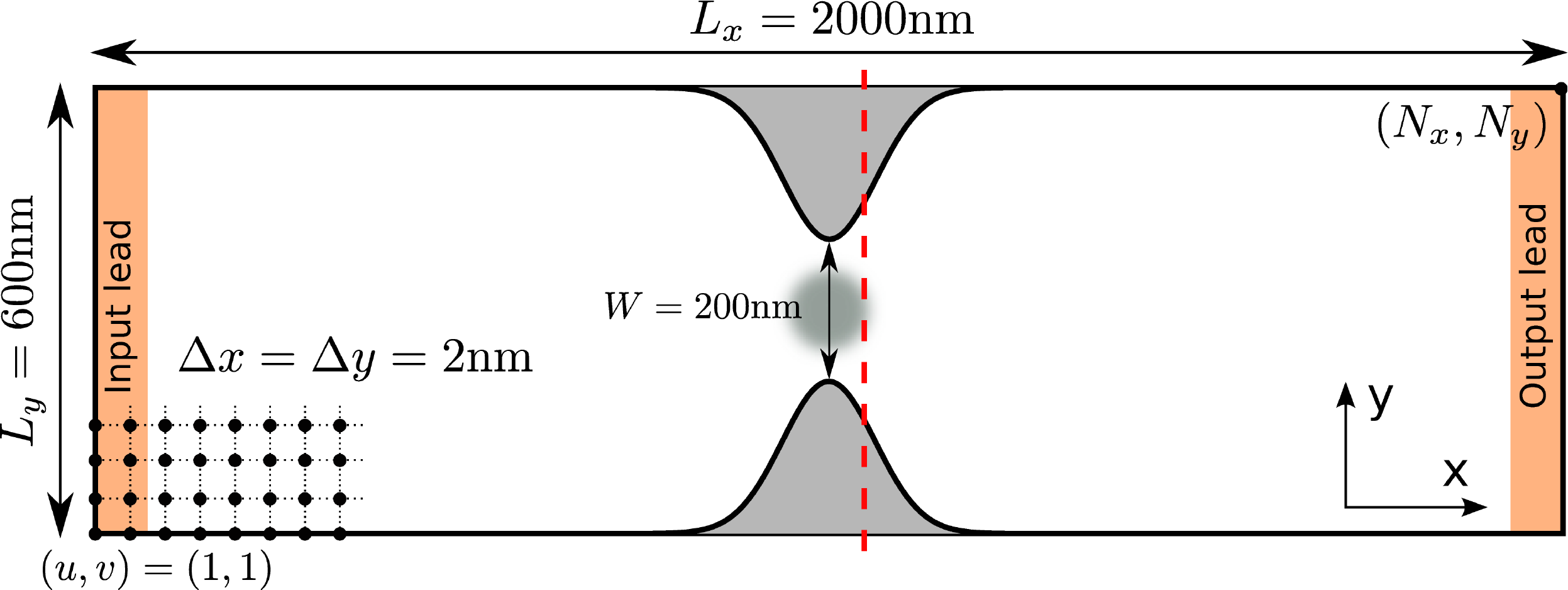}
\par\end{centering}
\caption{\label{fig:Sketch}(color online) Sketch of the QPC system considered
in this paper. The dots in the left corner show the finite difference
mesh used in numerical calculations. The vertical dashed red line shows the
path of the conductance scan discussed in the text. The gray area in the center of QPC
indicates a local potential maximum introduced to model the quantum Hall island.}
\end{figure*}

\begin{figure*}

\includegraphics[bb= 0 0 1500 500, width=15cm]{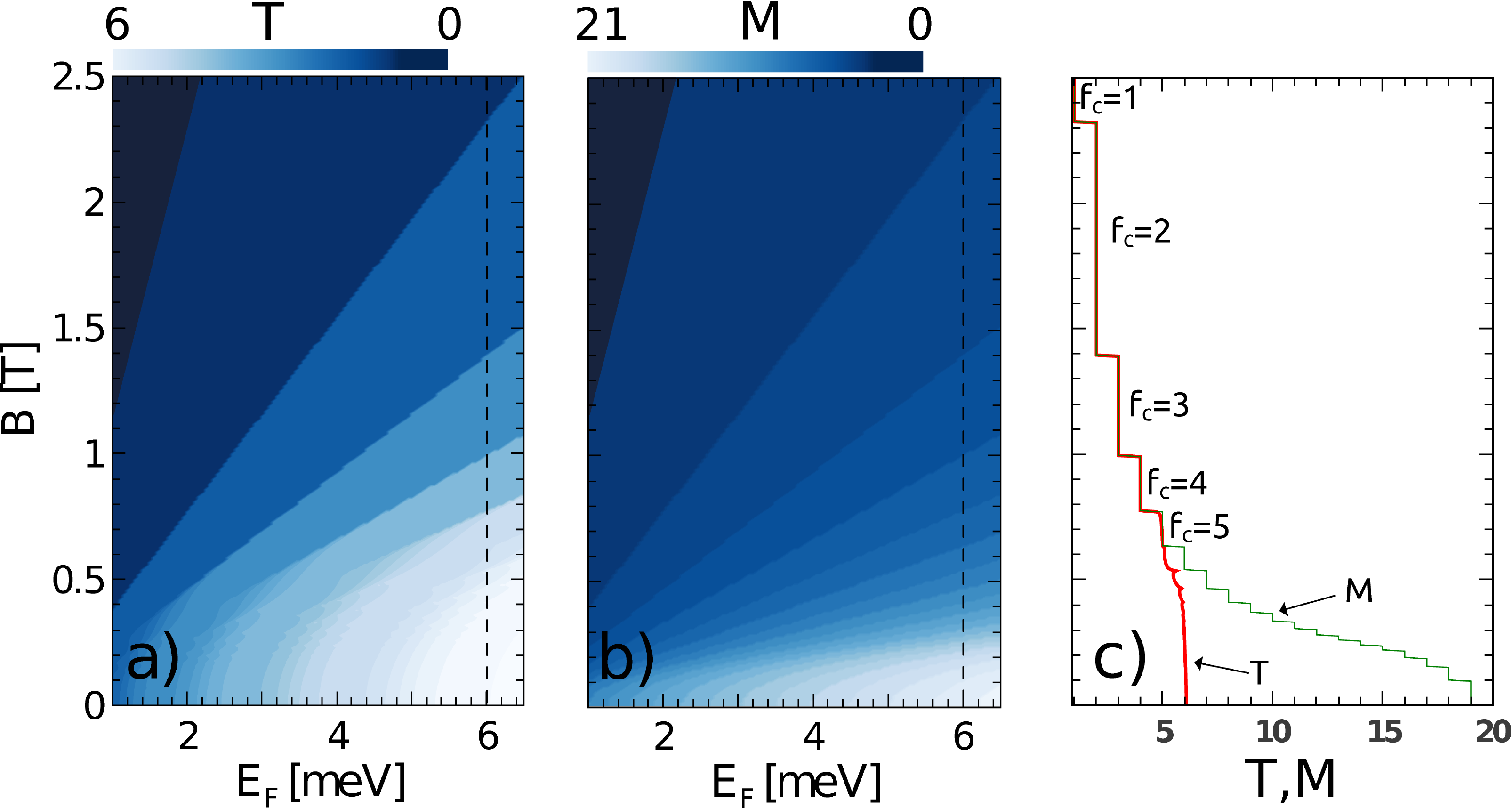}

\caption{\label{fig:czyste}(a) Transfer  probability through the QPC  summed over the incident
subbands [Eq. (7)] and (b) the number of transport modes  $M$ in the leads  as functions of
Fermi energy $E_{F}$ and the perpendicular magnetic field $B$.
The dashed lines on both plots show the energy value $E_{\mathrm{F}}=6\,\mathrm{meV}$
which is considered further in this paper. (c) Cross section of (a) and (b) for $E_{\mathrm{F}}=6\,\mathrm{meV}$.}
\end{figure*}

\begin{figure*}
\begin{centering}
\includegraphics[width=14cm]{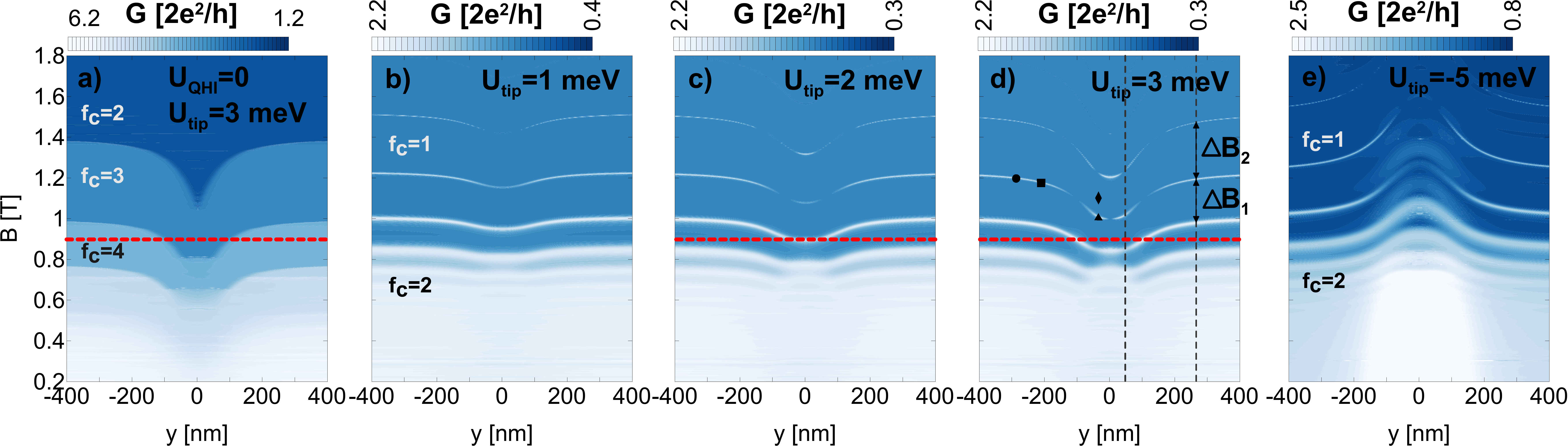}
\par\end{centering}
\caption{\label{fig:SGMy1} Conductance maps obtained for $B=0.9$ T and different values
of $U_{\mathrm{tip}}$ (a) without QHI, and (b-d) with QHI within QPC.}
\end{figure*}

We consider a wide channel with a narrowing that is presented in Fig. \ref{fig:Sketch}.
The channel has a width of $600$ nm that is reduced to 200 nm within the QPC.
We assume that the narrowing has a Gaussian shape (aspect ratio is preserved in Fig. \ref{fig:Sketch}).
The length of the computational box is $2000$ nm.

The Fermi level electron within the system is described
by a two-dimensional effective-mass Schr\"odinger equation
\begin{equation}
\left[\frac{1}{2m_{\mathrm{eff}}}(-i\hbar\boldsymbol{\nabla}+e\boldsymbol{A})^{2}+V(x,y)\right]\Psi(x,y)=E_{F}\Psi(x,y),\label{eq:main-schrod}
\end{equation}
 with the total potential $V(x,y)=V_c(x,y)+V_{\mathrm{tip}}(x,y)+V_{\mathrm{QHI}}(x,y)$,
 where $V_c$ is the confinement potential (we assume an infinite potential outside the channel and zero in the inside),
$V_\mathrm{tip}$ is the tip potential and $V_\mathrm{QHI}$ is the potential that models
the quantum  Hall island within the QPC. We consider a GaAs system with the effective electron band mass $m_{\mathrm{eff}}=0.067m_0$.
The spin Zeeman effect for  $B$ of the order of 1 T is still weak in GaAs and is neglected in the calculations.

The original potential of the tip as seen by the two-dimensional electron gas is of the Coulomb form.
This potential is screened by deformation of the gas.\cite{Szafran11}
In consequence the potential as seen by the Fermi level electrons is short-range.
Our previous Schr\"odinger-Poisson calculations \cite{Szafran11,chwiej13,kolasinski13} indicated that the effective tip potential
is close to Lorentzian with the width of the order of the distance between the tip and the electron gas.
Accordingly, in this paper we use the Lorentz model of the potential
 \begin{equation}
V_{\mathrm{tip}}(x,y)=\frac{U_{\mathrm{tip}}}{1+\left({\left(x-x_{\mathrm{tip}}\right)^{2}+\left(y-y_{\mathrm{tip}}\right)^{2}}\right)/{d_{\mathrm{tip}}^{2}}},\label{eq:lorentz}
\end{equation}
for the tip localized above point $(x_{\mathrm{tip}},y_{\mathrm{tip}})$.
We use $d_{\mathrm{tip}}=60$ nm for the width of the tip potential.
The potential of the quantum Hall island is also taken in the Lorentz form
 \begin{equation}
V_{\mathrm{QHI}}(x,y)=\frac{U_{\mathrm{QHI}}}{1+\left({\left(x-x_{\mathrm{QHI}}\right)^{2}+\left(y-y_{\mathrm{QHI}}\right)^{2}}\right)/{d_{\mathrm{QHI}}^{2}}},\label{eq:qhi}
\end{equation}
we assume that the QHI is located in the center of the constriction (see Fig. 1) with $x_\mathrm{QHI}=1000$ nm and $y_\mathrm{QHI}=0$.

We choose the Lorentz gauge $\boldsymbol{A}=(-By,0,0)$ for the uniform magnetic field applied perpendicular
to the plane of confinement.
The brown areas at the ends of the computational box in Fig. \ref{fig:Sketch} denote the asymptotic regions where
the boundary conditions are introduced.
The calculation method applied here is a variant of the one
used previously in Ref. \onlinecite{Szafran11}.
We use the gauge-invariant
kinetic-energy discretization,\cite{Governale} which leads to the following
finite difference equation
\begin{eqnarray}
\Psi_{u,v}\left(4t_{0}+V_{u,v}-E_{F}\right)+\Psi_{u+1,v}\left(-t_{0}C_{x}^{*}\right)\label{eq:shrod} \\+\Psi_{u-1,v}\left(-t_{0}C_{x}\right)
+\Psi_{u,v-1}\left(-t_{0}\right)+\Psi_{u,v+1}\left(-t_{0}\right) & = & 0,\nonumber
\end{eqnarray}
 where $C_{x}=e^{-i\frac{e}{\hbar}\Delta x A_{x}}$, and $t_{0}=1/(2m_{\mathrm{eff}}\Delta x^{2}$).
For the considered magnetic fields and Fermi wave vectors, convergent
results are obtained for $\Delta x=\Delta y=2$ {nm}.

Equation (\ref{eq:shrod}) defines a set of linear equations for
the wave function in the interior of the computational box.
The boundary conditions for the scattering problem are set in the following way.
In the leads far away from the QPC and the tip potential, the confinement potential is independent of $x$, i.e.  $V(x,y)\rightarrow V(y)$,
thus we can write the asymptotic Hamiltonian eigenfunctions as superpositions of plane waves
multiplied by transverse modes $\chi_k$.
Far away from the scattering region -- beyond the range of the evanescent modes --
the wave function takes the form \cite{Datta}
\begin{equation}
\Psi(x,y)=\sum_{k=1}^{M}a_{k}e^{ikx}\chi_{k}(y)+b_{k}e^{-ikx}\chi_{-k}(y),\label{eq:wave-asy}
\end{equation}
where $M$ is the number of subbands at the Fermi level, $k$ is the real wave vector, $\chi_{k}(y)$ [$\chi_{-k}(y)$] represents the $k$-th incoming (backscattered) transverse mode.
The transverse modes are
found by solving the eigenproblem for the homogeneous lead.\cite{Szafran11}
The coefficients $a_{k}$
and $b_{k}$ correspond to the incoming and the outgoing amplitudes,
respectively.
At the output lead we can write the solution in the form of superposition
of outgoing modes
\begin{equation}
\Psi(x,y)=\sum_{k=1}^{M}d_{k}e^{ikx}\chi_{k}(y),\label{eq:wave_out}
\end{equation}
 where $d_{k}$ is the amplitude of outgoing mode $\chi_{k}$.
The method applied in Refs. \onlinecite{Szafran11,chwiej13,kolasinski13} used an iterative scheme for
evaluation of the scattering amplitudes. Here we get rid of the iteration
employing the quantum transmitting boundary (QTB)
which
was originally developed \cite{Lent90,Lent94} for the finite element method.
Here we adapt QTB for the finite difference method.
 The details
of the present calculation are given in the Appendix.
 A similar procedure has been applied recently in Ref. \onlinecite{Huang12} for the current flow through ballistic nanodevices but in the absence of magnetic field.

After solution of the quantum scattering problem we evaluate the conductance by the Landauer-B\"uttiker formula
\begin{equation}
G=\frac{e^{2}}{h}T=\frac{e^{2}}{h}\sum_{i}^{M}T_{i}\label{eq:G},
\end{equation}
where $T_{i}$ is the transmission probability of the $i$'th mode incident from the input lead. The transmission $T_{i}$ for each incoming mode is calculated in
the following way.
For a given incoming mode $i$ we set the incoming amplitudes to $a_{k}=\delta_{ik}$,
where $k=(1,..,M)$. We solve the Schr\"odinger equation (\ref{eq:shrod}) with transmitting boundary conditions (see Appendix).
Then, we calculate incoming $b_{k}$ and outgoing $d_{k}$ amplitudes (Appendix).
Finally, the transmission probability is calculated from the probability current
fluxes
\begin{equation}
T_{i}=\sum_{k=1}^{M}\frac{\left|d_{k}\right|^{2}\sum_{v=1}^{N_{y}}\left|\chi_{k}(v)\right|^{2}\sin\left(\frac{e}{\hbar}\Delta x^{2}vB+k\Delta x\right)}{\sum_{v=1}^{N_{y}}\left|\chi_{i}(v)\right|^{2}\sin\left(\frac{e}{\hbar}\Delta x^{2}vB+k_{i}\Delta x\right)},\label{eq:Ti}
\end{equation}
where $k_{i}$ stands for the wave vector of the $i$'th incoming mode
and the expression $\sum_{v=1}^{N_{y}}\left|\chi_{k}(v)\right|^{2}\sin\left(\frac{e}{\hbar}\Delta x^{2}vB+k\Delta x\right)$
corresponds to the probability flux of a given mode (with $k>0$).

\section{Results}

\subsection{$G$ maps for a clean QPC}

Figure \ref{fig:czyste}(a) demonstrates the transfer probability summed over the subbands [Eq. (7)]
as a function of the magnetic field $B$ and the Fermi energy $E_{F}$.
The $T(B,E_F)$ function exhibits a step-like behavior with reduction of the number of the transport modes
with increasing $B$ or lowering $E_F$.
The results of Fig. \ref{fig:czyste}(a) are obtained from solution to the scattering
problem involving $M$ subbands in the leads -- see Fig. \ref{fig:czyste}(b) which appear at
the Fermi level for a given $B$. For the further discussion we choose Fermi energy equal to $6$ meV.

The $G$ map obtained for $B=0.9$ T for the clean QPC is presented in Fig. \ref{fig:SGMy1}(a).
The conductance is reduced from $f_c=4$ to $f_c=3$ when the tip approaches the area of the QPC.
Note, that when the tip is outside the QPC the $G$ map ignores its presence.
The flat minimum of $G$ within the QPC and the insensitiveness of the map to the position of the tip
when its outside the constriction is a characteristic feature of experimental maps obtained
in SGM imaging of the edge states in  QPCs -- see Ref. \onlinecite{edge2} of Ref. \onlinecite{paradiso} [Fig. 4].
Note that in contrast to the integer quantum Hall regime, for $B=0$ the $G$ maps collected from the outside of the QPC contain fine details 
with resolved branches \cite{qpc1,paradiso} as well as interference fringes \cite{qpc4} involving backscattering by the tip.
For high $B$ the backscattering is only allowed for the tip forming bridges between the conducting sample edges, hence
the flat region of the $G$ map outside the QPC.

\begin{figure*}
\begin{centering}
\includegraphics[bb= 0 0 2100 589, width=14cm]{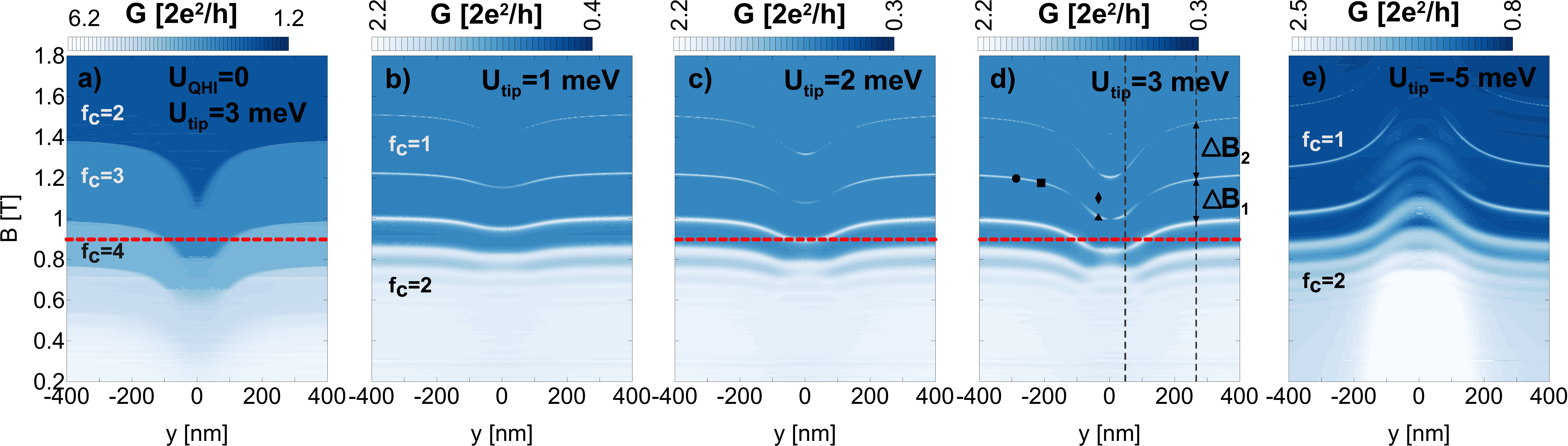}
\par\end{centering}
\caption{\label{fig:SkanYbezQHI}
Conductance for  $E_F=6$ meV
as a function of the $y$ position of the tip and the magnetic field.
The scans were performed along $x_{\mathrm{tip}}=1100$ nm line (see Fig. 1)
with $d_{\mathrm{tip}}=60$ nm for different values of the tip potential $U_{\mathrm{tip}}$ (Eq. \ref{eq:lorentz}).
(a) Scan of the QPC obtained with $U_{\mathrm{tip}}=3$meV and $U_{\mathrm{QHI}}=0$ (in the absence of QHI).
(b-e) Scans obtained for QHI present with: $U_{\mathrm{QHI}}=11$meV and $d_{\mathrm{QHI}}=40$nm (Eq. \ref{eq:qhi})
for $U_{\mathrm{tip}}=1,2,3$ and $-5$ meV.}
\end{figure*}

Figure \ref{fig:SkanYbezQHI}(a) shows the conductance for the tip scanning 
across the channel close to QPC along the red dashed line in Fig. 1 with varying magnetic field.
In the absence of the tip the $G(B)$ dependence has a step-like character [see Fig. \ref{fig:czyste}(a)]
in consistence with the experimental results of Ref. \onlinecite{edge2} [Fig. 2(f)] and the ones of Ref. \onlinecite{paradiso} [Fig. 4]
and Ref. \onlinecite{paradiso2} [Fig. 1(a)].
When the repulsive tip approaches the axis of the QPC ($y\simeq 0$) it enhances
the backscattering and induces shifts of $G$ steps to lower values of $B$.
The experimental result for the equivalent measurement with QHI inside the constriction [Fig. 2(a) of Ref. \onlinecite{martinsglowna}]
exhibits oscillations of $G$ instead of steps that appear in the result of Fig. \ref{fig:czyste}(a).

\begin{figure*}
\begin{centering}
\includegraphics[width=12cm]{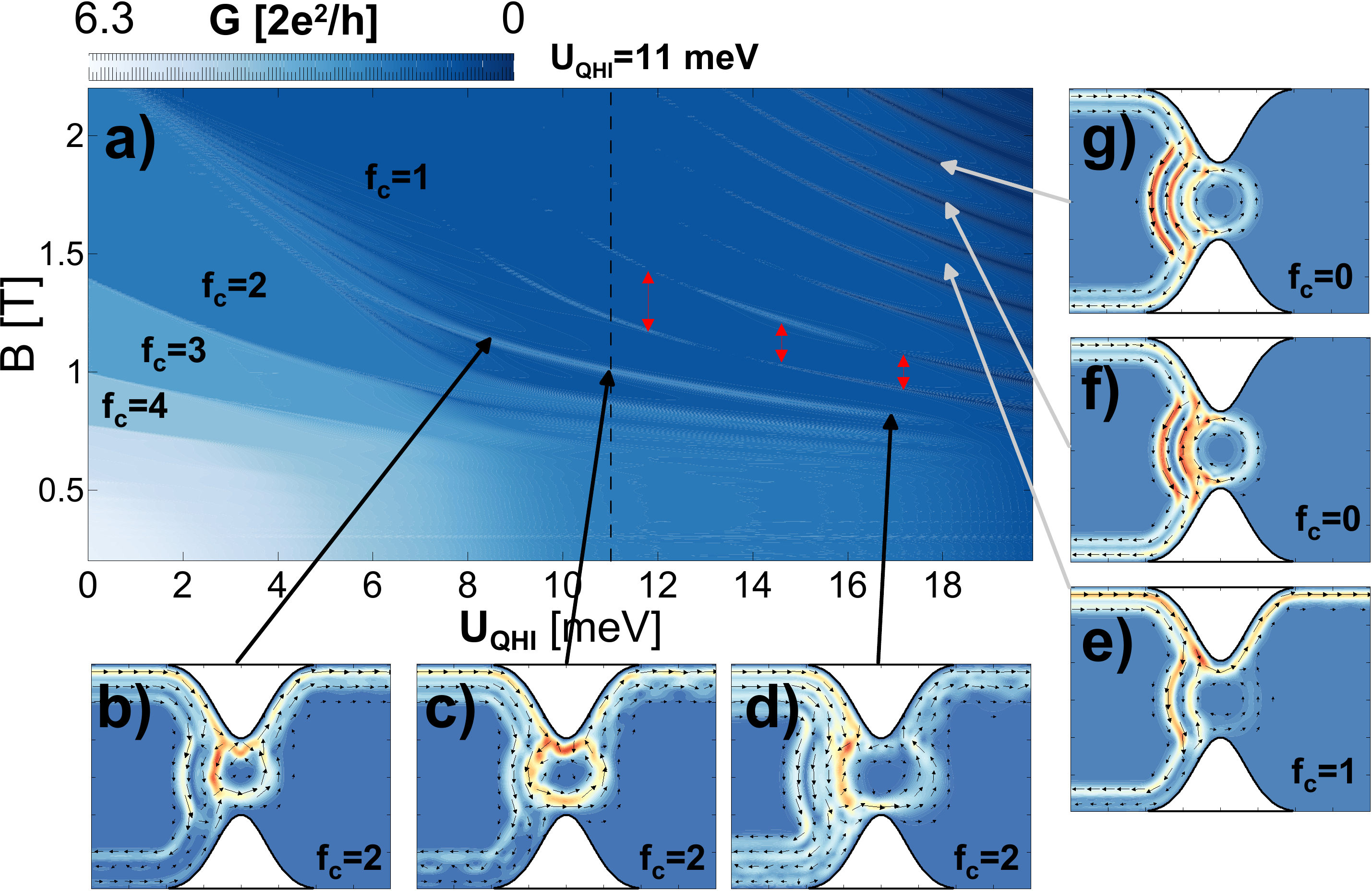}
\par\end{centering}
\caption{\label{fig:TodUqhi}(a) Transfer probability as a function of the
amplitude $U_{\mathrm{QHI}}$ and magnetic field for $d_{\mathrm{QHI}}=40$ nm and  the center of the potential maximum located at the center of QPC [see Fig.1].
The number of fully transparent subbands $f_c$ is given.
The vertical dashed line present the value of $U_{\mathrm{QHI}}$
which chose for further calculations. (b-g) Probability density current distribution (color scale shows the absolute value,
and vectors the orientation of the current)
for various points along resonances with locations indicated by arrows.
}
\end{figure*}

\begin{figure*}
\begin{centering}
\includegraphics[width=14cm]{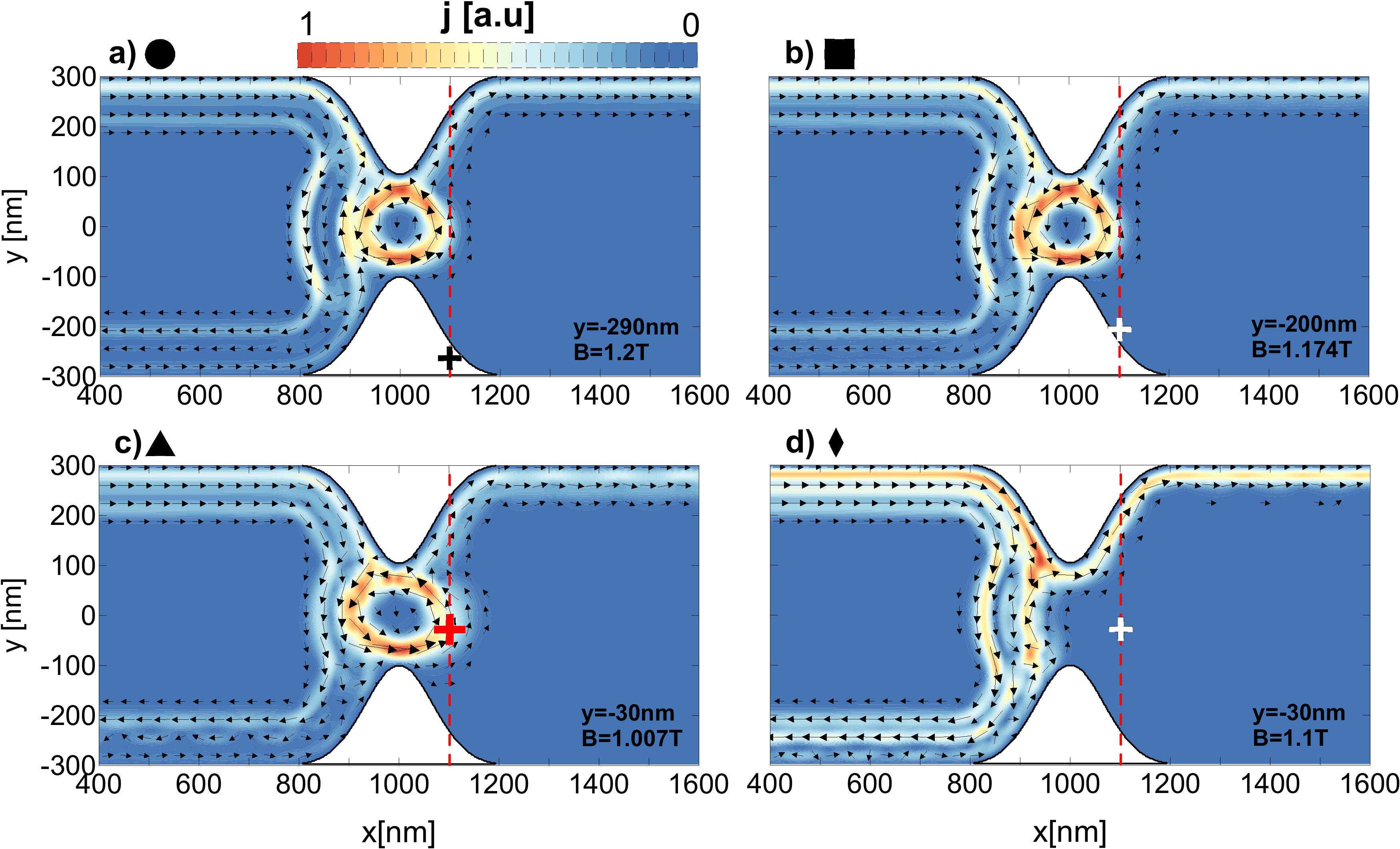}
\par\end{centering}
\caption{\label{fig:prady} (a-c) Probability density current distribution (color scale shows  the absolute value,
and the arrows -- the orientation of the current) corresponding to points
marked by symbols in the conductance scan of Fig \ref{fig:SkanYbezQHI}(d).
The position of the tip is marked by cross. }
\end{figure*}

\begin{figure}
\begin{centering}
\includegraphics[width=7cm]{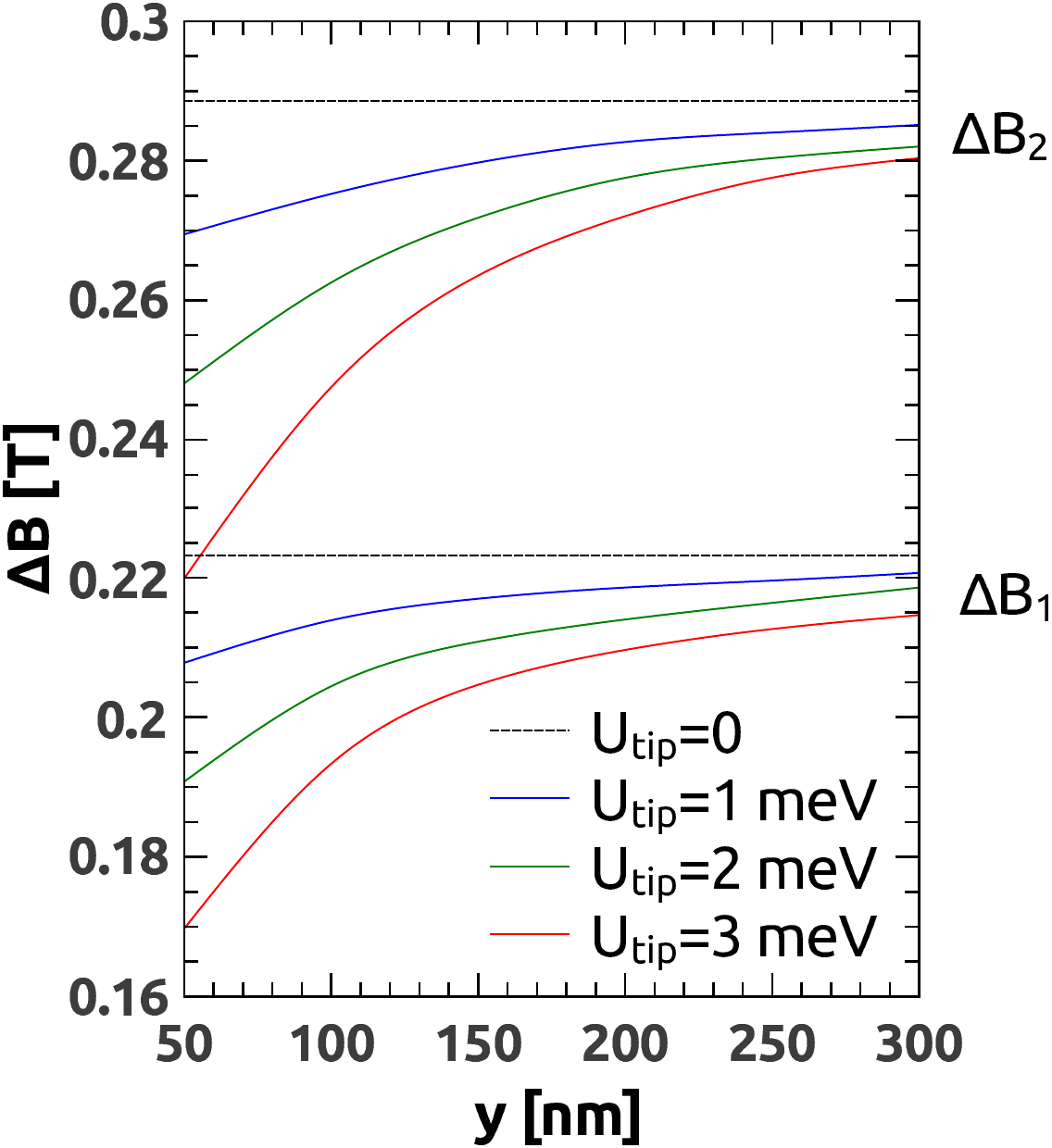}
\par\end{centering}
\caption{\label{fig:DbodBiUtip}$\Delta B$ dependence on the position $y$
of the tip for different values of $U_{\mathrm{tip}}$ as extracted from
Fig. \ref{fig:SkanYbezQHI}(b-d). For the definition of $\Delta B_1$ and $\Delta B_2$
see Fig. \ref{fig:SkanYbezQHI}(d). }
\end{figure}

\subsection{$G$ maps for the interferometer}
In search for the oscillatory behavior of conductance
in function of the tip position we have introduced a local potential maximum to the center of the QPC in order to simulate the quantum Hall island \cite{martinsglowna}
with radius $d_{QHI}=40$ nm [see Eq. (\ref{eq:qhi})].
The results for the transfer probability as a function of the magnetic field and the height
of the QHI potential perturbation are given in Fig. \ref{fig:TodUqhi}.
For low $B$ an increase of $U_{QHI}$ monotonically reduces the conductance.
However, at higher $B$ oscillations of conductance appear.
The period of these oscillations decreases with $U_{QHI}$ (see the red arrows),
which suggests that the resonances correspond to currents circulating around
the QHI with area increasing with the value of the local potential maximum.
The currents in the selected locations of the $(B,U_{QHI})$ diagram
were presented in Fig. \ref{fig:TodUqhi}(b-g).
In all the plots we find that the current approaches the QPC along
the upper edge and is partially backscattered to the lower one. The transmitted current
stays close to the upper edge. The area of the QHI is surrounded by an anticlockwise current
with the electron density pushed to the left of the current direction in consistence with
the classical Lorentz force orientation.\cite{szepl}
We find that the loops of current are fully developed for both $G$ resonances [Fig. \ref{fig:TodUqhi}(b-d)]
and antiresonances [\ref{fig:TodUqhi}(g,f)]. Outside the resonances and antiresonances the current
vortex is weak [Fig. \ref{fig:TodUqhi}(e)].
We conclude that the presence of a potential defect inside the QPC leads to formation of closed current loops
which is reentrant in function of $B$ for values
of the magnetic field which are more or less periodically spaced.
The current loops are coupled to the edge currents, hence the periodic features of conductance.
For the further discussion we fix $U_{QHI}=11$ meV for which  $f_c\ge 1$  for $B$ up to 2.2 T.

The $G$ maps obtained for $B=0.9$ T for the QHI present inside the QPC are displayed in Fig. \ref{fig:SGMy1}(b-d).
Instead of a central flat minimum of conductance found for the clean QPC [Fig. \ref{fig:SGMy1}(a)] a resonant ring localized around
the QHI defect is detected in agreement with the experimental results.\cite{martinsglowna}
The radius of the ring increases with $U_{tip}$.

Figures {\ref{fig:SkanYbezQHI}}(b-d) show scans of conductance along the line at a side of the QPC [see Fig. 1] for the increasing value of the tip potential.
Already for the tip outside the channel ($y$=400 nm), the conductance exhibits peaks which reappear nearly periodically as
functions of $B$ [Fig. \ref{fig:TodUqhi}].
 The repulsive tip changes the position of the resonances shifting them to lower magnetic field [Fig. {\ref{fig:SkanYbezQHI}}(b-d)]
 in accordance with the experiment.\cite{martinsglowna}.
An opposite shift is found for the attractive tip [Fig.{\ref{fig:SkanYbezQHI}}(e)].

In Fig. \ref{fig:prady} we plotted the current distribution for three points following the
resonance of Fig. \ref{fig:SkanYbezQHI}(d) that are marked by ($\circ,\square,\triangle$) and one point ($\diamond$) outside the
resonance [Fig. \ref{fig:SkanYbezQHI}(e)].
The resonances are related to the interference with current circulating
around the QHI. The repulsive tip potential increases the area encircled by the current when placed near the QHI [cf. Fig. \ref{fig:DbodBiUtip}(a) and Fig. \ref{fig:DbodBiUtip}(c)].
In consequence, we find that the period of the oscillations is reduced [see Fig. \ref{fig:DbodBiUtip}] when the tip
is near QPC.
Note, that for fixed $B$ the tip even when far from the center of the QPC destroys the resonant current loop
while moving along the straight line (note the radial form of $G$ map of Fig. 3).
In the discussed plots of Fig. 5(b-f)
the values of $B$ had to be changed to follow the resonance.
Moreover,  for the attractive tip
[Fig. \ref{fig:SkanYbezQHI}(e)] the spacing between the resonances increases when tip gets close to the center of the QPC, indicating
that the area encircled by the current is decreased.
The present calculations confirm the interpretation of Ref. \onlinecite{martinsglowna} that the reduced period of $G$ oscillations
for the repulsive tip is a signature of the presence of QHI inside the QPC.
Note, that the spacings between $B$ values -- already in the absence of the tip are not exactly equal (see Fig. \ref{fig:DbodBiUtip} for $U_{tip}=0$), which results from
the softness of the assumed QHI potential. For $B>0$ the current circulates counter-clockwise around the QHI [see Fig. \ref{fig:TodUqhi}(b-g)].
The Lorentz force pushes the electron density to the left of the current.
For higher magnetic field the shift of the electron density to the center of QHI is stronger, hence the reduced area of the loop [cf.  Fig. \ref{fig:TodUqhi}(b) and (d)]
and the increased Aharonov-Bohm period.

\begin{figure*}
\begin{centering}
\includegraphics[width=10cm]{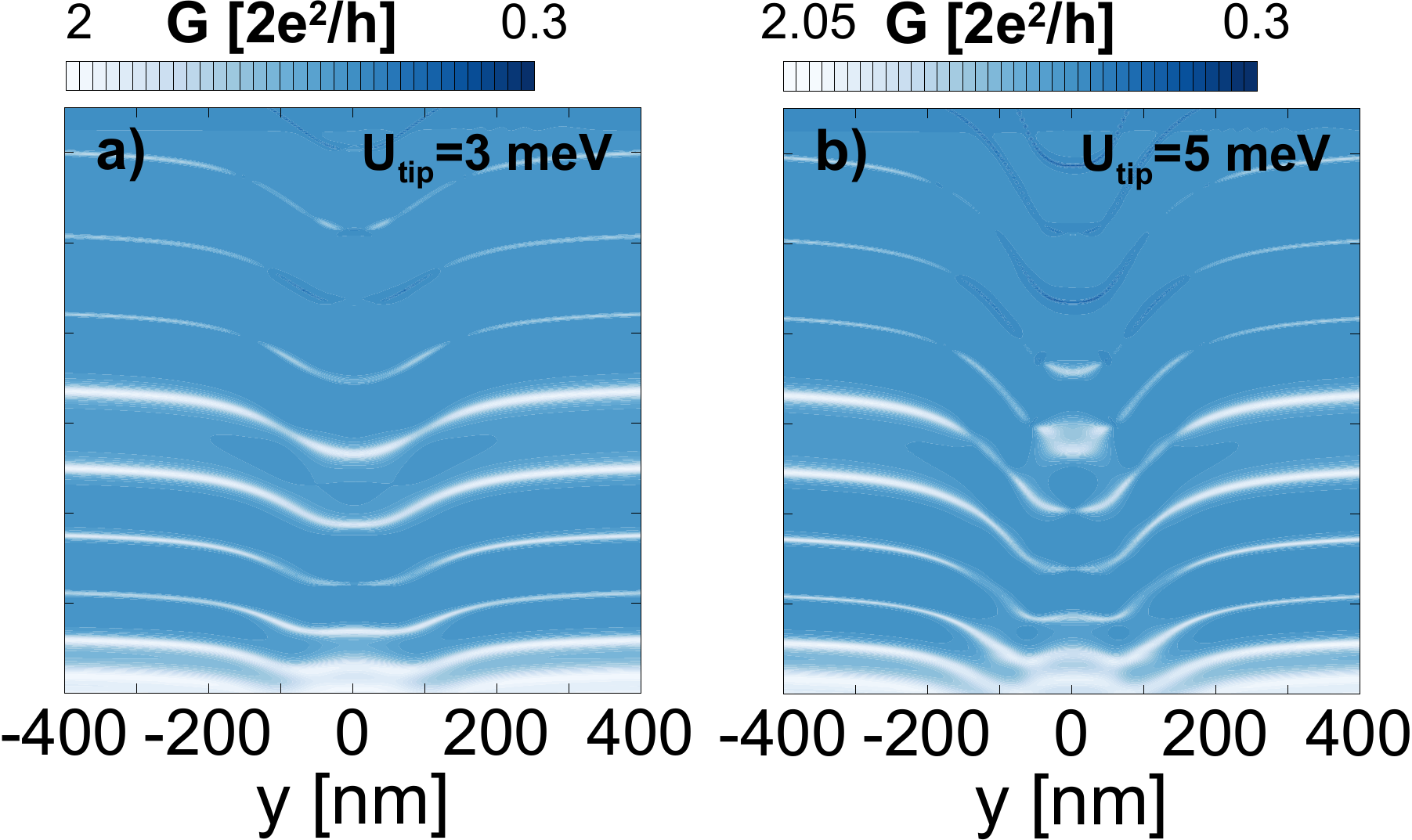}
\par\end{centering}
\caption{\label{fig:HardQHI}  Same as Fig. \ref{fig:SkanYbezQHI} only for the hard-wall profile of the QHI [Eq. (\ref{hw})].}
\end{figure*}

The change of the Aharonov-Bohm period with the presence of the tip results from superposition of the two potentials: the one of the defect forming
the QPC and the tip potential. We find that in order for this superposition to be effective in the modulation of the $B$ period the potential of the defect needs to be soft.
We have performed calculations for a hard-wall QHI potential simulated by
\begin{equation}
V_{\mathrm{QHI}}(x,y)=U_{\mathrm{QHI}}e^{\left(-\left[\left(\left(x-x_{\mathrm{QHI}}\right)^{2}+\left(y-y_{\mathrm{QHI}}\right)^{2}\right)/r_{\mathrm{QHI}}\right]^{8}\right)}.\label{hw}
\end{equation}
For the purpose of the hard-wall simulation we adopted $U_\mathrm{QHI}=10$ meV exceeding by 4 meV the Fermi energy, and $r_{QHI}=70$ nm.
The results for the scan along the path marked in Fig. 1 by the dashed line are given in Fig. \ref{fig:HardQHI}.
In contrast to the results with the soft QHI potential of Fig. \ref{fig:SkanYbezQHI}
we notice that i) in the absence of the tip ($y=400$ nm) the subsequent $G$ resonances at the $B$ scale are spaced by periods which change with the magnetic field
much more slowly than for the soft QHI defect and ii) the presence of the tip shifts the position of the resonances
but does not change their spacing significantly as it was the case for the soft QHI defect.

\section{Summary and Conclusions}
We have simulated scanning gate microscopy mapping of conductance of both a clean QPC and the one
turned into an electron interferometer by a local maximum of the potential landscape in the integer
quantum Hall regime.
We have solved the quantum scattering problem as given by the Schr\"odinger equation
using a direct finite difference approach with an implementation of the quantum transmitting boundary method.
We have described the stepwise reduction of conductance that is due to the tip for clean QPC as well as
formation of resonant current loops around the potential defect when introduced to the constriction.
We found that the repulsive tip reduces the period of the Aharonov-Bohm-like conductance oscillations for the interferometers.
We demonstrated that the periodicity of AB oscillations reacts to the tip only when the potential defect within QPC has a soft
character. The presented results for the conductance maps are consistent with the recent experimental results for both clean QPCs
and the electron interferometer.

\section*{Appendix}
This Appendix contains the details of the implementation of the quantum transmitting boundary method
for the finite difference approach.
After Refs. \onlinecite{Lent90,Lent94} we multiply both sides of Eq. (\ref{eq:wave-asy}) by a complex conjugate
of the m-th outgoing transverse mode $\chi_{-m}$, then we integrate
both sides along the channel, putting $x=0$
\begin{equation}
\left\langle \chi_{-m},\Psi\right\rangle =\sum_{k=1}^{M}a_{k}\left\langle \chi_{-m},\chi_{k}\right\rangle +b_{k}\left\langle \chi_{-m},\chi_{-k}\right\rangle ,\label{eq:leq}
\end{equation}
where $\left\langle A,B\right\rangle =\Delta x\sum_{j=1}^{N_{y}}A_{j}^{*}B_{j}$
is a standard inner product calculated for the transverse wave function across the channel, here written in finite difference formalism.
Equation (\ref{eq:leq}) can be written as system of $M$-th linear equations
for outgoing amplitudes $b_{k}$
\begin{equation}
\boldsymbol{v}=\boldsymbol{A}\boldsymbol{a}+\boldsymbol{S}^{-1}\boldsymbol{b},\label{eq:input_eq}
\end{equation}
 with
$v_{m}  =  \left\langle \chi_{-m},\Psi\right\rangle$,
$A_{mk}  =  \left\langle \chi_{-m},\chi_{k}\right\rangle$,
$S_{mk}^{-1}  =  \left\langle \chi_{-m},\chi_{-k}\right\rangle$,
$\boldsymbol{a}  =  (a_{1},a_{2},\ldots,a_{M})^T$, and
$\boldsymbol{b}  =  (b_{1},b_{2},\ldots,b_{M})^T$.

Note, that the different lateral modes $\chi_{k}$ are not orthogonal in presence of the external magnetic field,\cite{Datta} and thus $\boldsymbol{S^{-1}}$ and its inverse
matrix are not diagonal. We use Eq. (\ref{eq:input_eq}) to express
vector $\boldsymbol{b}$ in terms of the incident amplitudes
\[
\boldsymbol{b}=\boldsymbol{Sv-SA}\boldsymbol{a},
\]
with
\begin{equation}
b_{k}=\sum_{p=1}^{M}S_{kp}\left\{ \left\langle \chi_{-p},\Psi\right\rangle -\sum_{q}^{M}A_{pq}a_{q}\right\} .\label{eq:bk}
\end{equation}

In order to apply the boundary conditions for Eq. (\ref{eq:shrod}), we
calculate the derivative of Eq. (\ref{eq:wave-asy}) at $x=0$ ($u=1$)
using the standard central finite difference formula for first order
derivative
\begin{equation}
\left.\frac{\partial\Psi}{\partial x}\right|_{u=1}=\frac{\Psi_{2,j}-\Psi_{0,j}}{2\Delta x}=\sum_{k=1}^{M}\Delta_{k}a_{k}\chi_{k}(j)-\Delta_{k}b_{k}\chi_{-k}(j),\label{eq:dpsi}
\end{equation}
where
\[
\Delta_{k}=\frac{e^{ik\Delta x}-e^{-ik\Delta x}}{2\Delta x}=-\Delta_{-k}.
\]

Now we put the Eq. (\ref{eq:bk}) to the Eq. (\ref{eq:dpsi}), where
we also use the formula for inner product
\[
\left.\left\langle \chi_{-m},\Psi\right\rangle \right|_{u=1}=\Delta x\sum_{i=1}^{N_{y}}\chi_{-m}^{*}(j)\Psi_{1,i},
\]\begin{widetext}
 we get
\begin{eqnarray*}
\frac{\Psi_{2,j}-\Psi_{0,j}}{2\Delta x} & = & \sum_{k=1}^{M}\Delta_{k}\left[a_{k}\chi_{k}(j)-\chi_{-k}(j)\sum_{p=1}^{M}S_{kp}\left\{ \left\langle \chi_{-p},\Psi\right\rangle -\sum_{q=1}^{M}A_{pq}a_{q}\right\} \right],\nonumber \\
 & = & \sum_{k=1}^{M}\Delta_{k}a_{k}\chi_{k}(j)+\sum_{k=1}^{M}\Delta_{k}\chi_{-k}(j)\sum_{p=1}^{M}S_{kp}\sum_{q=1}^{M}A_{pq}a_{q}\nonumber \\
 & - & \Delta x\sum_{i=1}^{N_{y}}\Psi_{1,i}\cdot\sum_{k,p=1}^{M}\chi_{-p}^{*}(i)S_{kp}\chi_{-k}(j)\Delta_{k}=F_{j}-\Delta x\sum_{i=1}^{N_{y}}\alpha_{i}^{j}\Psi_{1,i},\label{eq:deriv-input}
\end{eqnarray*}
where
\begin{eqnarray*}
F_{j} & = & \sum_{k=1}^{M}\Delta_{k}a_{k}\chi_{k}(j)+\sum_{k=1}^{M}\Delta_{k}\chi_{-k}(j)\sum_{p,q=1}^{M}S_{kp}A_{pq}a_{q},
\end{eqnarray*}
\[
\alpha_{i}^{j}=\sum_{k,p=1}^{M}\chi_{-p}^{*}(i)S_{kp}\chi_{-k}(j)\Delta_{k}.
\]
 From (Eq. \ref{eq:deriv-input}) we get
\begin{eqnarray*}
\Psi_{0,j} & = & \Psi_{2,j}-2\Delta x\left(F_{j}-\Delta x\sum_{i=1}^{N_{y}}\alpha_{i}^{j}\Psi_{1,i}\right),
\end{eqnarray*}
which we put into the Schr\"odinger equation (\ref{eq:shrod}) in order
to apply boundary conditions for nodes with $u=1$
\begin{eqnarray*}
\Psi_{1,v}\left(4t_{0}+V_{1,v}-E_{F}\right)-2t_{0}\Re\left\{ C_{x}\right\} \Psi_{2,v} & -\nonumber \\
2t_{0}C_{x}\Delta x^{2}\sum_{i=1}^{N_{y}}\alpha_{i}^{v}\Psi_{1,i}-t_{0}\left(\Psi_{1,v-1}+\Psi_{1,v+1}\right) & = & -2t_{0}C_{x}\Delta xF_{v},\label{eq:shrod-1}
\end{eqnarray*}

Since $2t_{0}\Delta x^{2}=1/m_{\mathrm{eff}}$, we get
\begin{eqnarray*}
\Psi_{1,v}\left(4t_{0}+V_{1,v}-E_{F}-\frac{C_{x}}{m_{\mathrm{eff}}}\alpha_{v}^{v}\right)-2t_{0}\Re\left\{ C_{x}\right\} \Psi_{2,v} & +\nonumber \\
\Psi_{1,v+1}\left(-t_{0}-\frac{C_{x}}{m_{\mathrm{eff}}}\alpha_{v+1}^{v}\right)+\Psi_{1,v-1}\left(-t_{0}-\frac{C_{x}}{m_{\mathrm{eff}}}\alpha_{v-1}^{v}\right) & -\label{eq:bc_in}\\
\sum_{i\neq\{v-1,v,v+1\}}^{N_{y}}\frac{C_{x}}{m_{\mathrm{eff}}}\alpha_{i}^{v}\Psi_{1,i} & = & -2t_{0}\Delta xC_{x}F_{v},\nonumber
\end{eqnarray*}
 which is the final formula for the boundary condition for the input
lead. Using Eq. (\ref{eq:wave_out}) and choosing the coordinate frame in
which $x=0$ at nodes with $u=N_{x}$ we get
\begin{equation}
d_{k}=\sum_{p}D_{kp}\left\langle \chi_{p},\Psi\right\rangle ,\label{eq:dk}
\end{equation}
\[
D_{kp}^{-1}=\left\langle \chi_{k},\chi_{p}\right\rangle .
\]
 The same as for the input lead one can show that the boundary condition
at the output lead $u=N_{x}$ is given by formula
\begin{eqnarray*}
\Psi_{N_{x},v}\left(4t_{0}+V_{N_{x},v}-E_{F}-\frac{C_{x}^{*}}{m_{\mathrm{eff}}}\beta_{v}^{v}\right)-2t_{0}\Re\left\{ C_{x}\right\} \Psi_{N_{x}-1,v} & +\nonumber \\
\Psi_{N_{x},v+1}\left(-t_{0}-\frac{C_{x}^{*}}{m_{\mathrm{eff}}}\beta_{v+1}^{v}\right)+\Psi_{N_{x},v-1}\left(-t_{0}-\frac{C_{x}^{*}}{m_{\mathrm{eff}}}\beta_{v-1}^{v}\right) & -\label{eq:bc_out}\\
\sum_{i\neq\{v-1,v,v+1\}}^{N_{y}}\frac{C_{x}^{*}}{m_{\mathrm{eff}}}\beta_{i}^{v}\Psi_{N_{x},i} & = & 0,\nonumber
\end{eqnarray*}
with
\begin{eqnarray*}
\beta_{i}^{j} & = & \sum_{k,p=1}^{M}\chi_{p}^{*}(i)D_{kp}\chi_{k}(j)\Delta_{k}.
\end{eqnarray*}
\end{widetext}

Equations (\ref{eq:shrod}), (\ref{eq:bc_in}) and (\ref{eq:bc_out}) define
a set of $N_{x}N_{y}$ algebraic equations for unknown nodal values
of $\Psi_{u,v}$, which we solve using the LU method for sparse matrices \cite{SuperLU}.

\end{document}